\documentclass{appolb}
\usepackage{graphicx}
\usepackage{amssymb,amsfonts,amsmath}
\newcommand{\be}{\begin{equation}}
\newcommand{\ee}{\end{equation}}

\newcommand{\ba}{\begin{eqnarray}}
\newcommand{\ea}{\end{eqnarray}}

\begin{document}
\title{Chiral perturbation theory
in the environment with chiral imbalance
\thanks{Talk given on the conference "Excited QCD" (Schladming, Austria,Jan. 30- Feb. 3, 2019).  } 
}
\author{A.~A.~Andrianov$^{1,2}$, V.~A.~Andrianov$^{1}$, D.~Espriu$^2$
\address{$^1$ Faculty of Physics, Saint Petersburg State University, Universitetskaya nab.
7/9, Saint Petersburg 199034,
Russia}
\address{
$^2$ Departament de F\'isica Qu\`antica i Astrof\`isica and Institut de Ci\`encies del
Cosmos (ICCUB), Universitat de
Barcelona, Mart\`i Franqu\`es 1, 08028 Barcelona, Spain
}}
\maketitle
\begin{abstract}
We discuss the description of chiral imbalance imprint in pion matter. It is based on the Chiral Perturbation Theory supplemented by medium induced chiral chemical potential in the covariant derivative. Such an implementation of chiral imbalance recently was explored in linear sigma model inspired by QCD. The relationship between two sigma models is examined. A possible experimental detection of chiral imbalance in charged pion decays inside the fireball is pointed out.
\end{abstract}
\bigskip

\section{ Chiral Imbalance} The generation of a phase with local parity breaking (LPB) in strong interactions in central heavy ions collisions (HIC) has attracted much interest \cite{7,aaep} although the experimental evidences are still poor.
After a HIC the LPB  in the so-called fireball is possible as a result of the appearance in a dense and hot nuclear environment  a difference between the densities of the right- and left-handed chiral fermion fields (Chiral Imbalance). Such a chiral medium can be simulated by a chiral  chemical potential $\mu_5$. Adding to the QCD lagrangian the term  $\Delta{\mathcal
L}_q=\mu_5 q^\dagger \gamma_5 q \equiv \mu_5 \rho_5$, we open the way of accounting for non-trivial
topological fluctuations \cite{aaep} in the nuclear (quark) fireball, which are related to fluctuations of gluon fields. The behavior of various spectral characteristics for light scalar and pseudoscalar ($\sigma, \pi^a, a^a_0$) - mesons by means of the QCD motivated $\sigma$ model lagrangian was recently derived  for $SU_L (2) \times SU_R (2)$ flavor symmetry with an isosinglet chiral chemical potential \cite{eqcd16}.
It is shown, in particular,  that exotic decays of scalar mesons arise  due to mixing of $\pi$ and $a_0$ meson states in the presence of chiral imbalance. The pion electromagnetic formfactor obtains an unusual parity-odd supplement from the Wess-Zumino vertices which entails a photon polarization asymmetry of the pion polarizability in $\pi\gamma \to \pi\gamma$ process \cite{eqcd2017}.
However the structural constants of the model where taken as input parameters suitable to describe the light meson properties in a chiral medium. In this way there is no real predictability in predicting the hadron system response on chiral imbalance and reaching quantitative predictions requires a phenomenologically justified hadron dynamics. This is provided to some extent by Chiral Perturbation theory (ChPT) \cite{GL} with structural constants verified in low energy interactions of pseudoscalar mesons.

To follow this way one can reckon on the quark-hadron continuity ~\cite{contin} when passing through hadronization and for the detection of LPB in the hadron fireball
implement the chiral lagrangian model with a background 4-vector of axial chemical potential ~\cite{eqcd2017}, symmetric under
$ SU_{L}(N_{f})\times SU_{R}(N_{f})$, for $ u,d$-quarks ($N_{f}=2$).

Chiral imbalance can be associated with gradient density of isosinglet pseudoscalar condensate which can be formed
as a result of  large-scale, "long-lived" topological fluctuations
 of gluon fields
 in the fireball in central heavy ion collisions (see
~\cite{aaep} for details). To describe various effects of hadron
matter in a fireball with LPB, we must introduce the
chiral chemical potential ~\cite{aaep}.




\section{Chiral lagrangian with chiral chemical potential}
For the detection of LPB in the hadron
fireball one considers the Chiral Lagrangian for pions  describing mass spectra and decays of pseudoscalar mesons in the fireball carrying a chiral imbalance. The latter can be implemented with the help of softly broken chiral symmetry in QCD transmitted to hadron media, properly constructed long derivative
\begin{equation}
D_\nu \Longrightarrow  \bar D_\nu - i \{{\bf I}_q\mu_5 \delta_{0\nu}, \star
\}={\bf I}_q \partial_\nu -i A_\nu [Q_{em}, \star]- 2i{\bf I}_q\mu_5 \delta_{0\nu},\label{covderu}
\end{equation}
where $A_\mu, Q_{em}$ are the electromagnetic field and charge respectively.
The axial chemical potential
is introduced as a constant time component of an
isosinglet axial-vector field.

Further on, we skip the electromagnetic field. In the framework of large color number $N_c$ \cite{kais} the SU(3) Chiral Lagrangian in the strong interaction sector contains the following dim=2 operators \cite{kais},
\be
{\cal L}_2=\frac{F^2_0}{4} <- j_\mu j^\mu  +\chi^\dagger U+\chi U^\dagger>,
\label{L2lag}
\ee
where $<...>$ denotes the trace in flavor space, $j_\mu \equiv U^\dagger \partial_\mu U $, the chiral field $U=\exp(i \hat\pi/ F_0)$,\ $F_0 \simeq 92 MeV$, $\chi(x)=2B_0 s(x)$ and $M_\pi^2=2B_0 \hat m_{u,d}$,  the tree-level neutral pion mass. The constant $B_0$ is related to the chiral quark condensate $<\bar q q>$ as $ F_0^2 B_0 = − <\bar q q>$.
Taking now the covariant derivative in \eqref{covderu} it yields
\begin{equation}
{\cal L}_2 (\mu_5)={\cal L}_2(\mu_5=0)+\mu_5^2 N_f F_0^2.
\end{equation}
Herein we have used the identity for $U\in SU(n)$,$< j_\mu>=0$.
In the large $N_c$ approach the dim=4 operators \cite{kais} in the Chiral Lagrangian are given by
\ba
{\cal L}_4 =\bar L_3  <j_\mu j^\mu j_\nu j^\nu> + L_0 <j_\mu j_\nu j^\mu j^\nu> - \bar L_5 < j_\mu j^\mu (\chi^\dagger U+\chi U^\dagger)>,\label{L4lag}
\ea
where $L_0, \bar L_3, \bar L_5$ are bare low energy constants.
For SU(3) and SU(2) $<j_\mu>=0$ and there is the identity
\ba &&<j_\mu j_\nu j^\mu j^\nu> \\
&&=-2 <j_\mu j^\mu j_\nu j^\nu> + \frac12 <j_\mu j^\mu> <j_\nu j^\nu> +  <j_\mu j_\nu> <j^\mu j^\nu> ,\nonumber \label{id1}\ea
whereas for SU(2) there is one more identity
\be 2 <j_\mu j^\mu j_\nu j^\nu> = <j_\mu j^\mu> <j_\nu j^\nu> . \label{id2}\ee
Applying these identities one finds the four-derivative GL operators for the SU(3) chiral lagrangian
\be{\cal L}_4 = L_1 <j_\mu j^\mu> <j_\nu j^\nu> +  L_2 <j_\mu j_\nu> <j^\mu j^\nu> +L_3  <j_\mu j^\mu j_\nu j^\nu>\ee with
\be L_1=  \frac12 L_0;\ L_2= L_0;\ L_3 =\bar L_3 -2L_0.\ee
For SU(2) one has a further reduction of the dim=4 lagrangian,
\be {\cal L}_4 = \frac14 l_1 <j_\mu j^\mu> <j_\nu j^\nu> + \frac14 l_2 <j_\mu j_\nu> <j^\mu j^\nu> \label{final}\ee
with normalization so that
\be  l_1 = 2 L_0 +2 L_3,\  l_2 =  4 L_0,\  (l_1 +l_2) = 2L_3 + 6 L_0.\ee
We stress that this chain of transformations is valid only if $<j_\mu>=0$.

The response of the Chiral Lagrangian on chiral imbalance is derived with the help of the long derivative \eqref{covderu} applied to the lagrangian \eqref{L4lag},
\ba
&&\Delta {\cal L}_4 (\mu_5) = -4 \mu_5^2\{( l_1 +l_2) <j_\mu j^\mu> + (\l_1+ l_2 )\mathfrak{}<j^0 j^0>\}\nonumber\\&& =
-4 \mu_5^2\{ 2(l_1 + l_2) <j^0 j^0> -(l_1+l_2) <j_k j_k> \}.
\ea
We notice that this result is drastically different from what one could obtain from the final lagrangian \eqref{final}. This is because the identities \eqref{id1} and \eqref{id2} are violated if $<j_\mu>\not=0$
 The mass term affected by chiral imbalance comes out from a dim=4 GL vertex \eqref{L4lag},
 \be
 \Delta {\cal L}_4 (\mu_5)= l_4\mu_5^2 <\chi^\dagger U+\chi U^\dagger>;\quad l_4 = 4\bar L_5.
 \ee
The above modifications change the decay constants differently the coefficients in dispersion law in energy $p^0$ and three-momentum $|\vec p|$ for the mass shell  as well as modify the mass term for pions (all together  the inverse propagator of pions),
\ba
&&K(p^2) (\mu_5) = (F_0^2 + 32 \mu^2_5 (l_1 + l_2)) p_0^2\nonumber\\ && -(F_0^2 +16 \mu^2_5 (l_1+l_2) )|\vec p|^2 - (F^2_0 + 4l_4\mu_5^2) m_\pi^2\Big] \to 0.
\ea
The empirical values of the SU(2) GL SU(2) constants \cite{GL} normalized at the RG scale $\mu \simeq M_\pi \simeq 140 MeV, \ \log\Big(m_\pi/\mu\Big)\simeq 0$ are,
\ba
&&l_1^r = (-0.4 \pm 0.6)\times 10^{-3};\ l_2^r = (8.6 \pm 0.2) \times 10^{-3};\nonumber\\&& l_1^r + l_2^r = (8.2 \pm 0.8) \times 10^{-3};\
l_4^r =  (2,64 \pm 0.01)\times 10^{-2}.
\ea
Hence the distortion of mass shell can be detected in decays of charged pions when the effective pion mass approaches to muon mass which may happen if $\mu_5 \sim F_\pi$ .
\section{Linear sigma model for light pions and scalar mesons in the presence of chiral imbalance}
Let us compare these constants with those ones estimated from the sigma model build in \cite{eqcd2017,efflag}.
The sigma model was build with realization of SU(2) chiral symmetry to describe pions and isosinglen and isotriplet scalar mesons.
Its lagrangian reads
\begin{eqnarray}
{L}&=& N_c \Big\{\frac{1}{4}\,<(D_{\mu}H\,(D^{\mu}H)^{\dagger}>
+\frac{B_0}{2}\,  <m(H\,+\,H^{\dagger}>
+\frac{M^{2}}{2}\,<HH^{\dagger}>
\nonumber \\
\,&-&\frac{\lambda_{1}}{2}\,<(HH^{\dagger})^{2}>
-\frac{\lambda_{2}}{4}\, <(HH^{\dagger})>^{2}
+\frac{c}{2}\,(\det H +\,\det H^{\dagger})\Big\},
\label{lagr_sigma}
\end{eqnarray}
where $H = \xi\,\Sigma\,\xi$ is an operator for meson fields, $N_c$ is a number of colours,
\(m\) is an average mass of current $u,d$ quarks,
\(M\) is a "tachyonic" mass generating the spontaneous breaking of chiral symmetry,
\(B_0, c,\lambda_{1},\lambda_{2}\) are real constants.

The matrix \(\Sigma\) includes the singlet scalar meson \(\sigma\), its vacuum average \(v\) and the isotriplet of scalar mesons \( a^{0}_{0},a^{-}_{0},a^{+}_{0}\), the details see in ~\cite{eqcd2017,efflag}. The covariant derivative of $H$ including the chiral chemical potential $\mu_5$ is defined in \eqref{covderu}.
The operator \(\mathbf \xi\) realizes a nonlinear representation (see \eqref{L2lag}) of the chiral group $SU(2)_L \times SU(2)_R$, namely,  $\xi^2 = U$.
From spectral characteristics of scalar mesons in vacuum one fixes the lagrangian parameters,
\(\lambda_{1}=16.4850\), \(\lambda_{2}=-13.1313\), \(c=-4.46874\times10^4 {MeV}^2\), \(B_0  =1.61594\times10^5 {MeV}^2\).

The change of pion coupling constant $F_0 \simeq v$ is determined by potential parameters as compared to the ChPT definition,
\be
{\Delta F_\pi^2\over\mu_5^2} = \frac{1}{\lambda_1 +\lambda_2}\approx 0.3\quad \mbox{\it vs}\quad 32  (l_1 + l_2) \approx 0.26.
\ee
It is a quite satisfactory correspondence. It  makes it plausible that the above low energy GL operators with chiral imbalance are saturated by the isotriplet scalar meson exchange.

 Analogously, in the rest frame using the pion mass correction,\\ $ m^2_\pi(\mu_5)F^2_\pi(\mu_5) \simeq 2m_q B_0 F_\pi(\mu_5) $ it is easy to find the estimation for
 \be l_4 \approx 2,64 \times 10^{-2} \quad \mbox{\it vs}\quad \frac{1}{8(\lambda_1 +\lambda_2)}\approx 3.8\times 10^{-2}, \ee
 wherefrom one can also guess the relation $4(l_1 + l_2)  \sim l_4$ following from the linear sigma model.

 \section{Conclusions and prospects}
 We have found a convincing correspondence between the ChPT prognosis and the linear sigma model predictions for the dispersion law of distorted mass shell for pions under influence of chiral imbalance.
 The immediate consequence of such a distortion is a suppression of charged pion decay into muon and neutrino with increasing of chiral chemical potential. In the limiting case this decay channel may even become closed.

A manifestation for
LPB in the presence of chiral imbalance in the sector of  $\rho$ and $\omega$ vector
mesons\cite{eqcd16} can also happen and in this case the Chern-Simons interaction plays the main role.  It turns out\cite{aaep} that the spectrum of massive vector mesons splits into
three components with different polarizations having
different effective masses $m^2_{V,+} < m^2_{V,L}< m^2_{V,-}$.

We draw also attention to the recent proposal to measure the photon
polarization asymmetry in $\pi \gamma$ scattering \cite{harada,eqcd2017} as a way to detect LPB due
to chiral imbalance. It happens in the ChPT including electromagnetic fields due to the Wess-Zumino-Witten operators.

The independent check of our estimates could be done by lattice computation (see \cite{braguta} for inspiration).

{\bf Acknowledgements}\\
We express our gratitude to Angel G\'omez Nicola for stimulating discussion of how to implement chiral imbalance in ChPT. The funding for this work was provided by the Spanish MINECO under project
MDM-2014-0369 of ICCUB (Unidad de Excelencia `Maria de Maeztu'), grant FPA2016-76005-C2-1-P and grant 2014-SGR-104(Generalitat de Catalunya),
  by Grant RFBR 18-02-00264 and by  SPbSU travel grant Id: 36273714 (A.A.).


\begin{thebibliography}{99}
\bibitem{7} D. Kharzeev,{\it  Phys. Lett. B} {\bf 633}, 260 (2006)\\
 D. E. Kharzeev, L. D. McLerran, and H. J.
Warringa,{\it  Nucl. Phys. A}{\bf  803}, 227 (2008)\\
 K. Fukushima,
D. E. Kharzeev, and H. J. Warringa,{\it  Phys. Rev. D}{\bf  78}, 074033 (2008)
\bibitem{aaep} A. A. Andrianov, V. A. Andrianov, D. Espriu and X. Planells, {\it  Phys.
    Lett. B} {\bf 710}, 230 (2012);\ {\it Phys. Rev. D}{\bf 90},034024 (2014).
\bibitem{eqcd16}
 A.~A.~Andrianov, V.~A.~Andrianov and D.~Espriu,
  {\it Acta Phys.\ Polon.\ Supp. }  {\bf 9}, 515 (2016),{\it ibid.}{\bf 10}, 977 (2017)
  \bibitem{eqcd2017}
 A.~A.~Andrianov, V.~A.~Andrianov, D.~Espriu, A. V. Iakubovich, and A. E. Putilova,
 {\it Phys. Part. Nucl. Lett.}{\bf 15},357 (2018);\ {\it EPJ Web of Conferences}{\bf 191}, 05014 (2018).
 \bibitem{GL} J. Gasser and H. Leutwyler,{\it Annals Phys.}{\bf 158}, 142 (1984);\\
J. Bijnens and G. Ecker,{\it Annu. Rev. Nucl. Part. Sci.}{\bf  64:6.1–6.26} (2014)
\bibitem{contin} K. Fukushima and T. Hatsuda, {\it Rept. Prog. Phys. }{\bf 74}, 014001 (2011)
\bibitem{kais}
 R. Kaiser and H. Leutwyler, Eur. Phys. J. C 17, 623 (2000).
\bibitem{efflag} A. A. Andrianov, D. Espriu, and X. Planells, {\it  Eur. Phys. J. C} {\bf
    73}, 2294 (2013).
\bibitem{harada} M.~Kawaguchi, M.~Harada, S.~Matsuzaki and R.~Ouyang, {\it Phys. Rev. C}
      {\bf 95}, 065204 (2017).
\bibitem{braguta} V.~V.~Braguta et al.,  {\it  Phys. Rev. D} {\bf 93}, 034509 (2016);\quad
    V.~V.~Braguta, A.~Yu.~Kotov, {\it Phys. Rev. D}{\bf 93}, 105025 (2016).
\end{thebibliography}
\end{document}